\documentclass[prc,preprint,showpacs,tightenlines,%
                floatfix,amsfonts,nofootinbib]{revtex4}

\usepackage{bm}
\usepackage{amssymb}

\begin{document}

%
%
%
%
\newcommand{\covdev}{{\widetilde \partial}}
\newcommand{\finalnewpage}{\newpage}
\newcommand{\lagrang}{{\cal L}}
\newcommand{\newg}{{\skew2\overline g}_\rho}
\newcommand{\Nbar}{\skew3\overline N \mkern2mu}
\newcommand{\stroke}[1]{\mbox{{$#1$}{$\!\!\!\slash \,$}}}
\newcommand{\Ximinus}{\xi \frac{\tau_3}{2} \xi^\dagger%
                     - \xi^\dagger \frac{\tau_3}{2} \xi}
\newcommand{\Xiplus}{\xi \frac{\tau_3}{2} \xi^\dagger%
                     + \xi^\dagger \frac{\tau_3}{2} \xi}
\newcommand{\Xiaminus}{\xi \frac{\tau_a}{2} \xi^\dagger%
                     - \xi^\dagger \frac{\tau_a}{2} \xi}
\newcommand{\Xiaplus}{\xi \frac{\tau_a}{2} \xi^\dagger%
                     + \xi^\dagger \frac{\tau_a}{2} \xi}

%
%
\def\dthree#1{\intback{\rm d}^3\intback #1}
\def\dthreex{\dthree{x}}
\def\fpi{f_{\pi}}
\def\gammafive{\gamma^5}
\def\gammafivel{\gamma_5}
\def\gammamu{\gamma^{\mu}}
\def\gammamul{\gamma_{\mu}}
\def\intback{\kern-.1em}
\def\kfermi{k_{\sssize {\rm F}}}    
\def\mn{{\mu\nu}}
\def\psibar{\overline\psi}
\def\sigmamunu{\sigma^\mn}
\def\sigmamunul{\sigma_\mn}
\def\Tr{\mathop{\rm Tr}\nolimits}

\let\dsize=\displaystyle
\let\tsize=\textstyle
\let\ssize=\scriptstyle
\let\sssize\scriptscriptstyle
%
%
%


\newcommand{\beq}{\begin{equation}}
\newcommand{\eeq}{\end{equation}}
\newcommand{\beqa}{\begin{eqnarray}}
\newcommand{\eeqa}{\end{eqnarray}}

\def\Inthelimit#1{\lower1.9ex\vbox{\hbox{$\
   \buildrel{\hbox{\Large \rightarrowfill}}\over{\scriptstyle{#1}}\ $}}}

\title{Electromagnetic Interactions in a\\
       Chiral Effective Lagrangian for Nuclei}

\author{Brian D. Serot}\email{serot@indiana.edu}
\affiliation{Department of Physics and Nuclear Theory Center
             Indiana University, Bloomington, IN\ \ 47405}

%
\author{\null}
\noaffiliation

%
\date{\today\\[20pt]}

\begin{abstract}
\hfill\\[-20pt]     

Electromagnetic (EM) interactions are incorporated in a recently
proposed effective field theory of the nuclear many-body problem.
Earlier work with this effective theory exhibited EM couplings that
are correct only to lowest order in both the pion fields and the
electric charge.  The Lorentz-invariant effective field theory
contains nucleons, pions, isoscalar scalar $(\sigma )$ and vector
$(\omega )$ fields, and isovector vector $(\rho )$ fields. The
theory exhibits a nonlinear realization of $SU(2)_L \times SU(2)_R$
chiral symmetry and has three desirable features: it uses the same
degrees of freedom to describe the currents and the
strong-interaction dynamics, it satisfies the symmetries of the
underlying QCD, and its parameters can be calibrated using
strong-interaction phenomena, like hadron scattering or the
empirical properties of finite nuclei. It has been verified that for
normal nuclear systems, the effective lagrangian can be expanded
systematically in powers of the meson fields (and their derivatives)
and can be truncated reliably after the first few orders.  The
complete EM lagrangian arising from minimal substitution is derived
and shown to possess the residual chiral symmetry of massless,
two-flavor QCD with EM interactions. The uniqueness of the minimal
EM current is proved, and the properties of the isovector vector and
axial-vector currents are discussed, generalizing earlier work. The
residual chiral symmetry is maintained in additional (non-minimal)
EM couplings expressed as a derivative expansion and in implementing
vector meson dominance. The role of chiral anomalies in the EM
lagrangian is briefly discussed.
\end{abstract}

\smallskip
\pacs{24.10.Cn, 24.10.Jv, 12.39.Fe, 12.40.Vv.}

\maketitle

\section{Introduction}
\label{sec:intro}

Covariant meson--baryon effective field theories of the nuclear
many-body problem (often called \emph{quantum hadrodynamics} or QHD)
have been known for many years to provide a realistic description of
the bulk properties of nuclear matter and heavy nuclei.  (See
Refs.~\cite{SW86,Reinhard89,Gambhir90,BDS92,Ring96,SW97} for a
review.) Recently, a QHD effective field theory (EFT) has been
proposed \cite{FST97,FSp00,FSL00,EvRev00,LNP641} that includes all
of the relevant symmetries of the underlying QCD. In particular, the
spontaneously broken $SU(2)_L \times SU(2)_R$ chiral symmetry is
realized nonlinearly. The motivation for this EFT and illustrations
of some calculated results are discussed in
Refs.~\cite{SW97,FST97,HUERTAS02,HUERTASwk,%
HUERTAS04,MCINTIRE04,MCINTIRE05,JDW04}. This QHD EFT has also been
applied to a discussion of the isovector axial-vector current in
nuclei \cite{AXC}.

One advantage of this QHD EFT is that the electromagnetic (EM)
structure of the nucleon and pion can be introduced directly into
the lagrangian using a derivative expansion and vector meson
dominance (VMD)
\cite{FST97,SAKURAI60,GMZ,GMSW62,KLZ,BERNSTEIN68,SAKURAI69}, in
addition to the usual ``minimal'' couplings to the photon field.
Nevertheless, the discussion of the EM couplings in
Ref.~\cite{FST97} was brief and included only the terms needed for
the calculations in that work. In particular, nearly all of the
exhibited EM couplings are correct \emph{only to lowest order in
both the pion fields and the electromagnetic charge}.  The purpose
of this work is to illustrate the full structure of the
corresponding EM lagrangian, including all terms arising from the
introduction of EM gauge-covariant derivatives, and to extend and
clarify (and correct) the discussion in Ref.~\cite{FST97}.  This
full lagrangian will serve as the basis for a calculation of the
Lorentz-covariant, one- and two-nucleon amplitudes for both electron
scattering and pion photoproduction in the nuclear many-body
problem, which will be the subject of a forthcoming publication
\cite{CEH}. It also serves as a launching point for extending the
lagrangian to include the $\Delta$ resonance
\cite{TANGELLIS96,TANGELLIS97,TANGELLIS98} as an explicit degree of
freedom in the EM interactions.

This QHD EFT has three desirable features: (1) It uses the same
degrees of freedom to describe the currents and the
strong-interaction dynamics; (2) It respects the same internal
symmetries, both discrete and continuous, as the underlying QCD
(before and after electromagnetic interactions are included); and
(3) Its parameters can be calibrated using strong-interaction
phenomena, like $\pi$N scattering and the properties of finite
nuclei (as opposed to electroweak interactions with nuclei).

There are many, many papers in the literature that discuss EM
interactions in the context of chiral, effective, hadronic field
theory. (For example, see
Refs.~\cite{KLZ,GG69,BANDO85,GSS88,EGPdR,Urech95,BKM95,NR96,Scherer03}.)
This earlier work has focused almost entirely on EM and radiative
meson decays, EM contributions to meson masses, and the EM structure
of mesons and nucleons, the latter being carried out primarily in
the heavy-baryon formalism \cite{JM91}.

In the present work, we focus on EM interactions in the nuclear
many-body problem. We derive the relevant interactions for a
Lorentz-invariant QHD lagrangian that contains nucleons and $\pi$,
$\sigma$, $\omega$, and $\rho$ mesons \cite{FST97,SW97}. This
lagrangian has a \emph{linear} realization of the $SU(2)_V$ isospin
symmetry and a \emph{nonlinear} realization of the spontaneously
broken $SU(2)_L \times SU(2)_R$ chiral symmetry (when the pion mass
is zero).  It was shown in
Refs.~\cite{FST97,FSp00,LNP641,FML96,FRIAR97} that by using Georgi's
naive dimensional analysis (NDA) \cite{GEORGI93} and the assumption
of \emph{naturalness} (namely, that all appropriately defined,
dimensionless couplings are of order unity), it is possible to
truncate the lagrangian at terms involving only a few powers of the
meson fields and their derivatives, at least for systems at normal
nuclear densities \cite{MS96}.  It was also shown that a mean-field
approximation to the lagrangian could be interpreted in terms of
density functional theory \cite{Dre90,SW97,SW01,LNP641,EDFT}, so
that calibrating the parameters to observed bulk and single-particle
nuclear properties (approximately) incorporates many-body effects
that go beyond Dirac--Hartree theory.  Explicit calculations of
closed-shell nuclei provided such a calibration and verified the
naturalness assumption \cite{FSp00,FML96}. {\em This approach
therefore embodies the three desirable features needed for a
description of electromagnetic interactions in the nuclear many-body
problem.}

We will work in the chiral limit, since the structure of terms
involving explicit chiral symmetry breaking is well known
\cite{GL84,Wei96}, and these terms do not change our currents.  It
is apparent from the lagrangian of two-flavor, massless QCD that
when the photon is introduced with the familiar, local $U(1)_Q$
charge symmetry,\footnote{%
For $u$ and $d$ quarks in an isospinor $\psi$, the coupling to the
photon $A^\mu$ is ${-e} A^\mu \psibar \gamma_\mu Q \psi = {-e} A^\mu
( \psibar_L \gamma_\mu Q \psi_L + \psibar_R \gamma_\mu Q \psi_R )$,
where $\psi_{R,L} \equiv \frac{1}{2}\, (1 \pm \gammafivel ) \psi$,
and the charge matrix is $Q = \frac{1}{2} (\frac{1}{3} + \tau_3 )$.
Since the chiral symmetry is realized with global rotation matrices,
this interaction is invariant under independent left- and
right-handed rotations about the third isospin axis.} a residual,
global, chiral symmetry remains, which involves left- and
right-handed isospin rotations about the ``3'' axis: $U(1)_{L_3}
\times U(1)_{R_3} \times U(1)_B$. (There is also the usual global
phase symmetry associated with baryon number $B$.) This residual
global symmetry \emph{must also be present} in the low-energy QHD
EFT. We explicitly exhibit this symmetry of our EM lagrangian and
discuss how our results are equivalent to the more familiar
procedure that uses external sources \cite{GL84}.  [For a lagrangian
with only $SU(2)$ isospin and chiral symmetry, as opposed to
$SU(3)$, there is no technical advantage to the more formal
approach.]

Moreover, we omit the fourth-order pion--pion and pion--nucleon
lagrangian ${\cal L}_4$, whose structure (and electromagnetic
interactions) are also well known \cite{Wei96}, since it has not
(yet) been relevant in nuclear many-body calculations. Finally, we
consider only terms with explicit photon fields. (For a discussion
of virtual-photon counterterms, see Refs.~\cite{Urech95,NR96}.)

It is important to note that in our QHD EFT, only pions and nucleons
(the hadronically stable particles) can appear on external lines
with \emph{timelike} four-momenta. The heavy non-Goldstone bosons
appear only on internal lines (with \emph{spacelike} four-momenta)
and allow us to parametrize the medium- and short-range parts of the
nucleon--nucleon interaction, as well as the electromagnetic form
factors of the hadrons using VMD~\cite{TANG95,FST97}. The heavy
bosons are also convenient degrees of freedom for describing
nonvanishing expectation values of bilinear nucleon operators, like
$\Nbar N$ and $\Nbar \gamma^{\mu} N$, which are important in nuclear
many-body systems~\cite{SW86,SW97}. Vacuum-loop contributions
involving heavy bosons (and nucleons) \emph{are not to be
calculated}, since they depend on short-range effects that should be
absorbed in the counterterms \cite{HU00}.

The remainder of this paper is organized as follows. Section
\ref{sec:EFT} briefly reviews the EFT lagrangian of
Refs.~\cite{FST97,AXC}. Section \ref{sec:minimal} defines our
decomposition of the EM lagrangian, introduces the EM
gauge-covariant derivatives, and derives the (unique) form of the
minimal EM current.  The residual symmetries of the EM lagrangian
and the divergences of the vector and axial-vector currents are also
discussed, as is the relationship to the external-field approach.
Section \ref{sec:nonminimal} introduces non-minimal EM interactions
in a gradient expansion that is used to describe the pion and
nucleon EM structure, and Sec.~\ref{sec:VMD} discusses the VMD
contributions. Section \ref{sec:anomalies} contains a brief
discussion of the anomalous EM couplings (and other abnormal-parity
interaction terms). Section \ref{sec:summary} is a summary.

\section{Effective Field Theory Lagrangian}
\label{sec:EFT}

The effective field theory (EFT) lagrangian considered in the
present work was proposed in Ref.~\cite{FST97}. As discussed in that
paper, the nonlinear chiral lagrangian can be organized in
increasing powers of the fields and their derivatives. To each
interaction term we assign an index
\begin{equation}
 \nu \equiv d + {n \over 2} + b \ ,
\end{equation}
where $d$ is the number of derivatives, $n$ is the number of nucleon
fields, and $b$ is the number of non-Goldstone boson fields in the
interaction term. Derivatives on the nucleon fields are not counted
in $d$ because they will typically introduce powers of the nucleon
mass $M$, which will not lead to small expansion
parameters~\cite{FST97}.

It was shown in Refs.~\cite{FST96,FST97} that for finite-density
applications at and below nuclear matter equilibrium density, one
can truncate the effective lagrangian at terms with $\nu \leq 4$. It
was also argued that by making suitable definitions of the nucleon
and meson fields, it is possible to write the lagrangian in a
``canonical'' form containing familiar noninteracting terms for all
fields, Yukawa couplings between the nucleon and meson fields, and
nonlinear meson interactions~\cite{FHT01}. See
Refs.~\cite{SW97,FST97} for a more complete discussion.

If we keep terms with $\nu \leq 4$, the chirally invariant
lagrangian
can be written as\footnote{%
We use the conventions of Refs.~\protect\cite{SW86,FST97,SW97,AXC}.}
\begin{eqnarray}
{\cal L}_{\mathrm{EFT}} & = & {\cal L}_N  + {\cal L}_4
        + {\cal L}_M  \nonumber \\[7pt]
 & = & \Nbar \left( i {\gamma}^{\mu} \left[ {\partial}_{\mu} + i v_{\mu}
+ i g_{\rho} {\rho}_{\mu} + i g_v V_{\mu} \right] + {g_A}\mkern2mu
  {\gamma}^{\mu} {\gamma}_{5} a_{\mu} - M + g_s \phi \right) N
        \nonumber \\[4pt]
 & & \quad
{} -  { {f_{\rho} g_{\rho}} \over {4 M} } \Nbar {\rho}_{\mu \nu}
{\sigma}^{\mu \nu} N - { {f_{v} g_{v}} \over {4 M} } \Nbar {V}_{\mu
\nu} {\sigma}^{\mu \nu} N - { { {\kappa}_{\pi} } \over M } \Nbar
{v}_{\mu \nu} {\sigma}^{\mu \nu} N + { {4 {\beta}_{\pi}} \over M }
\Nbar N \, {\rm Tr}
\left( a_{\mu} a^{\mu} \right) \nonumber \\[4pt]
 & & \quad
 {} + {\cal L}_4
+  { 1 \over 2 } \, {\partial}_{\mu} \phi \, {\partial}^{\mu} \phi +
{ 1 \over 4 } f^2_{\pi} \, {\rm Tr} \left({\partial}_{\mu} U
{\partial}^{\mu} U^{\dagger} \right) - { 1 \over 2 } \, {\rm Tr}
      \left( {\rho}_{\mu \nu} {\rho}^{\mu \nu} \right)
- { 1 \over 4 }\, {V}_{\mu \nu} {V}^{\mu \nu}
\nonumber \\[4pt]
 & &\quad
 {} - g_{\rho \pi \pi} { {2 f^2_{\pi}} \over { m^2_{\rho} } } \,
{\rm Tr} \left( {\rho}_{\mu \nu} {v}^{\mu \nu} \right) + { 1 \over 2
} \left( 1 + {\eta}_1 { {g_s \phi} \over M } + {{\eta}_2 \over 2} {
{g^2_s {\phi}^2} \over {M^2} } \right) m^2_v V_{\mu} V^{\mu}
 + { 1 \over {4!} } \,{\zeta}_0 g_v^2 {\left( V_{\mu} V^{\mu} \right)}^2
\nonumber \\[4pt]
 & &\quad
+ \left( 1 + {\eta}_{\rho} \, { {g_s \phi} \over M } \right)
m^2_{\rho} \, {\rm Tr} \left( {\rho}_{\mu} {\rho}^{\mu} \right) -
\left( { 1 \over 2 } + { {{\kappa}_3 } \over {3!} } { {g_s \phi}
\over M } + { {{\kappa}_4 } \over {4!} } { {g^2_s {\phi}^2} \over
{M^2} }\right) m^2_s {\phi}^2 , \label{eq:eft-lagrangian}
\end{eqnarray}
where the nucleon, pion, sigma, omega, and rho fields are denoted by
$N$, $ \bm{\pi}$, $\phi$, $V_\mu$, and $\rho_\mu \equiv { 1 \over 2
} \, \bm{\tau  \! \cdot \! \rho}_{\mu}$, respectively, with
$V_{\mu\nu} \equiv \partial_\mu V_\nu -
\partial_\nu V_\mu$, and ${\sigma}^{\mu \nu} \equiv {i \over 2}
[{\gamma}^{\mu}, {\gamma}^{\nu}]$. The trace ``Tr'' is in the $2
\times 2$ isospin space. The pion field enters through the
combinations
\begin{eqnarray}
U & \equiv & \exp(i \bm{\tau  \! \cdot \! \pi} / f_\pi ) \ , \qquad
\qquad
 \xi  \equiv  \exp(i \bm{\tau  \! \cdot \! \pi} / 2 f_\pi )
 \ , \label{eq:Upi} \\[5pt]
a_{\mu} &\equiv & - {i \over 2} \left( {\xi}^{\dag} {\partial}_{\mu}
{\xi} -
\xi {\partial}_{\mu} {\xi}^{\dag} \right) \ , \label{eq:a-mu}
                 \\[5pt]
v_{\mu} &\equiv&  - {i \over 2} \left( {\xi}^{\dag} {\partial}_{\mu}
{\xi} +
\xi {\partial}_{\mu} {\xi}^{\dag} \right) \ , \label{eq:v-mu}
                 \\[5pt]
v_{\mu \nu} &\equiv & \partial_\mu v_\nu - \partial_\nu v_\mu +
i[v_\mu ,
v_\nu ] = - i [a_{\mu},a_{\nu}] \ . \label{eq:v-mu-nu}
%
\end{eqnarray}
The rho meson enters through the covariant field tensor
\begin{equation}
{\rho}_{\mu \nu}=D_{\mu} {\rho}_{\nu} - D_{\nu} {\rho}_{\mu} +i \,
\newg [{\rho}_{\mu}, {\rho}_{\nu}] \ , \label{eq:rhofieldtensor}
\end{equation}
where the chirally covariant derivative is defined by
\begin{equation}
D_{\mu} {\rho}_{\nu} \equiv {\partial}_{\mu}{\rho}_{\nu} + i
[v_{\mu}, {\rho}_{\nu}] \ , \label{eq:rhoderiv}
\end{equation}
and $\newg$ is a free parameter~\cite{SW97,FST97}. ${\cal L}_4$
contains $\pi\pi$ and $\pi$N interactions of order $\nu = 4$ that
will not be considered further here. (These interactions are exactly
the same as in chiral perturbation theory
\cite{GSS88,TANGELLIS98,Bira99}.) Numerically small $\nu = 4$ terms
proportional to $\phi^2 \,{\rm Tr} \left( {\rho}_{\mu} {\rho}^{\mu}
\right)$ and $V_\mu V^\mu \,{\rm Tr} \left( {\rho}_{\mu}
{\rho}^{\mu} \right)$ have been omitted, although they have been
considered in Refs.~\cite{HP01a,HP01b}.

This EFT lagrangian provides a consistent framework for explicitly
calculating the two-body exchange currents originating from
meson--nucleon interactions in nuclei. According to naive
dimensional analysis (NDA), all of the coupling parameters are
written in dimensionless form and should be of order unity, if the
theory obeys naturalness; this has been verified for the parameters
that are relevant for mean-field nuclear structure
calculations~\cite{FST97,FSp00,FML96}. Moreover, all of the
constants entering the lagrangian~(\ref{eq:eft-lagrangian}) are
assumed to be determined from calibrations to nuclear and nucleon
structure data, hadronic decays, and $\pi$N scattering
observables~\cite{FST97,TANGELLIS98}.

The lagrangian of Eq.~(\ref{eq:eft-lagrangian}) exhibits a nonlinear
realization of $SU(2)_L \times SU(2)_R$ chiral symmetry
\cite{CWZ69,CCWZ69}. The transformation properties of the various
field combinations have been discussed many times and will not be
repeated here. (See, for example, Refs.~\cite{FST97,LNP641}.)  Under
arbitrary global transformations with matrices $L \in SU(2)_L$ and
$R \in SU(2)_R$, the fields are rotated by the local, so-called
``compensating-field'' matrix $h(x) \in SU(2)_V$, where $SU(2)_V$ is
the unbroken vector subgroup of $SU(2)_L \times SU(2)_R$.  The
matrix $h(x)$ becomes constant only for global $SU(2)_V$ (i.e.,
isospin) transformations, in which case $L = R = h$.

The full global symmetry $SU(2)_L \times SU(2)_R \times U(1)_B$
implies a conserved baryon current
\begin{equation}
B^\mu = \Nbar \gammamu N
         \label{eq:baryoncurr}
\end{equation}
and conserved isovector vector $(\mathbf{V}^\mu )$ and axial-vector
$(\mathbf{A}^\mu )$ currents, which can be determined using
Noether's theorem \cite{GML60,AXC}. The isovector currents are given
in the Appendix.  Noether's theorem also implies that these currents
are conserved only for fields that satisfy the Euler--Lagrange
equations.  The corresponding charges $Q^a$ and $Q^a_5$ are
constants of the motion and satisfy the familiar chiral charge
algebra \cite{AXC}.

\section{Minimal Electromagnetic Couplings}
\label{sec:minimal}

The EM interactions will be incorporated by adding to ${\cal
L}_{\mathrm{EFT}}$ of Eq.~(\ref{eq:eft-lagrangian}) the following
lagrangian:
\begin{equation}
\lagrang_\mathrm{EM} = \lagrang^\mathrm{min}_\mathrm{EM}
                        + \lagrang^\mathrm{had}_\mathrm{EM}
                        + \lagrang^\mathrm{vmd}_\mathrm{EM}
                        + \lagrang^\mathrm{anom}_\mathrm{EM} \ .
\label{eq:LEMfull}
\end{equation}
The four contributions describe, respectively,

\begin{itemize}
\item
$\lagrang^\mathrm{min}_\mathrm{EM}$: terms arising from minimal
substitution, obtained by replacing ordinary derivatives in ${\cal
L}_{\mathrm{EFT}}$ with EM gauge-covariant derivatives (these terms
are necessary);
\item
$\lagrang^\mathrm{had}_\mathrm{EM}$: non-minimal terms in a
derivative expansion, which will serve to describe some of the
hadronic EM structure;
\item
$\lagrang^\mathrm{vmd}_\mathrm{EM}$: VMD terms that contain the
coupling of the photon to neutral vector mesons (and pions);
\item
$\lagrang^\mathrm{anom}_\mathrm{EM}$: EM terms associated with
chiral anomalies, which describe, among other things, mesonic decays
like $\pi^0 \to \gamma\gamma$.
\end{itemize}

In the present section, we will be concerned with the minimal EM
couplings, which take the form
\begin{equation}
\lagrang^\mathrm{min}_\mathrm{EM} =
        -\frac{1}{4} F_{\mu\nu} F^{\mu\nu} - e A_\mu J^\mu_\mathrm{min}
          + \lagrang^\mathrm{min}_{e^2} \ ,
\label{eq:LEMmin}
\end{equation}
where $F_{\mu\nu} = \partial_\mu A_\nu -\partial_\nu A_\mu$ is the
usual EM field tensor.  The terms of $O(e^2)$ contain two factors of
the photon field $A^\mu$.  Possible terms of higher order in $e$
vanish due to the antisymmetry of the field tensors ${v}_{\mu \nu}$
and ${\rho}_{\mu \nu}$.  As we will show, the minimal current
$J^\mu_\mathrm{min}$ is conserved through $O(e^0)$.

To include the EM interactions, we elevate a subgroup of the full
global symmetry group to the status of a \emph{local} symmetry. This
necessitates the introduction of a massless gauge field $A_\mu$ and
the corresponding gauge-covariant derivatives of the matter fields.
The local $U(1)_Q$ symmetry is described by a one-parameter group
characterized by a generator (``electric charge'')
\begin{equation}
Q = \frac{1}{2} B + T_3 \ , \label{eq:Qdef}
\end{equation}
with $T_3 = Q_3$ the third component of isospin and $B$ the baryon
number.

Under $U(1)_Q$, the EM field $A_\mu$ transforms in the familiar way
\begin{equation}
A_\mu \to A_\mu - \frac{1}{e}\, \partial_\mu \alpha (x) \ .
\end{equation}
The pion, rho, and nucleon fields transform under a local $U(1)_Q$
rotation in the same fashion as noted earlier, with $L$, $R$, and
$h$ set equal to
\begin{equation}
q(x) \equiv \exp \left[ i \alpha (x) \left( \frac{B+ \tau_3}{2}
            \right) \right] \ ,
\label{eq:qdef}
\end{equation}
where $B=0$ for the pion and rho, and $B=1$ for the nucleon. Thus $N
\to qN$, $\xi \to q \xi q^\dagger$, $\rho_\mu \to q \rho_\mu
q^\dagger$, etc.

The EM interactions explicitly break the symmetry of ${\cal
L}_{\mathrm{EFT}}$.  The lagrangian can be made EM gauge invariant
by replacing the ordinary derivatives with the gauge-covariant
derivatives \cite{FST97}
\begin{eqnarray}
\covdev_\mu N &\!\!\equiv\!\!& \left[ \partial_\mu + \frac{i}{2}\, e
            A_\mu (1 + \tau_3 ) \right] N \ ,
            \label{eq:Ncovdev} \\[5pt]
\covdev_\mu U &\!\!\equiv\!\!& \partial_\mu U + i e A_\mu \left[
           \frac{\tau_3}{2}, U \right] \ ,
           \label{eq:Ucovdev} \\[5pt]
\covdev_\mu \xi &\!\!\equiv\!\!& \partial_\mu \xi + i e A_\mu \left[
         \frac{\tau_3}{2}, \xi \right] \ ,
         \label{eq:xicovdev} \\[5pt]
\covdev_\mu \rho_\nu &\!\!\equiv\!\!& \partial_\mu \rho_\nu + i e
         A_\mu \left[ \frac{\tau_3}{2}, \rho_\nu \right] \ ,
         \label{eq:rhocovdev}
\end{eqnarray}
and similarly for the adjoint fields. We will consistently use a
\emph{tilde} to distinguish EM gauge-covariant derivatives from
ordinary derivatives. As a result of the preceding definitions, the
axial and vector pion fields become
\begin{eqnarray}
{\widetilde a}_\mu &\!\! = \!\!& a_\mu + \frac{1}{2}\, e A_\mu
    \left(
    \xi^\dagger \left[ \frac{\tau_3}{2}, \xi \right] - \xi \left[
    \frac{\tau_3}{2}, \xi^\dagger \right] \right)
    = a_\mu +
    \frac{1}{2}\, e A_\mu \left( \xi^\dagger \frac{\tau_3}{2} \xi
    - \xi \frac{\tau_3}{2} \xi^\dagger \right) \ ,
    \label{eq:atwiddle} \\[5pt]
{\widetilde v}_\mu &\!\! = \!\!& v_\mu + \frac{1}{2}\, e A_\mu
    \left(
    \xi^\dagger \left[ \frac{\tau_3}{2}, \xi \right] + \xi \left[
    \frac{\tau_3}{2}, \xi^\dagger \right] \right)
    = v_\mu +
    \frac{1}{2}\, e A_\mu \left( \xi^\dagger \frac{\tau_3}{2}
    \xi + \xi \frac{\tau_3}{2} \xi^\dagger - \tau_3 \right) \ .
    \label{eq:vtwiddle}
\end{eqnarray}
[Note that Eq.~(\ref{eq:atwiddle}) differs by a minus sign from the
expression in Eq.~(32) of Ref.~\cite{FST97}.] The chirally covariant
derivative in Eq.~(\ref{eq:rhoderiv}) becomes
\begin{equation}
{\widetilde D}_\mu \rho_\nu = D_\mu \rho_\nu + \frac{i}{2}\, e A_\mu
    \left[
    \left( \Xiplus \right) , \rho_\nu \right] \ ,
    \label{eq:covDrho}
\end{equation}
and the pion field tensor becomes
\begin{equation}
{\widetilde v}_{\mu\nu} = v_{\mu\nu} + \frac{i}{2}\, e A_\mu \left[
    \left( \Ximinus \right) , a_\nu \right] - \frac{i}{2}\, e A_\nu
    \left[ \left( \Ximinus \right) , a_\mu \right] \ .
    \label{eq:covvmunu}
\end{equation}
It is straightforward to verify that the quantities in
Eqs.~(\ref{eq:Ncovdev}) through (\ref{eq:covvmunu}) all transform
\emph{homogeneously} under $U(1)_Q$, e.g., $\covdev_\mu N \to q
\covdev_\mu N$, $\covdev_\mu \xi \to q (\covdev_\mu \xi )
q^\dagger$, ${\widetilde v}_{\mu\nu} \to q {\widetilde v}_{\mu\nu}
q^\dagger$, etc.

Using these gauge-covariant derivatives and functions, it is now a
straightforward (but tedious) exercise to gauge the original EFT
lagrangian and to express the result in the form ${\cal
L}_{\mathrm{EFT}} + \lagrang^\mathrm{min}_\mathrm{EM}$
[Eq.~(\ref{eq:LEMmin})], with
\begin{eqnarray}
J^\mu_\mathrm{min} &\!\!=\!\!&
    -i\, \frac{\fpi^2}{4} \Tr [\tau_3 (U \partial^\mu U^\dagger + U^\dagger
                             \partial^\mu U )]
    + \frac{1}{2}\, \Nbar \gammamu \left( 1 + \Xiplus \right) N
               \nonumber\\[5pt]
& & \quad
    {}+ \frac{1}{2}\, g_A \Nbar \gammamu\gammafivel \left( \Ximinus \right) N
    + i\, \frac{\kappa_\pi}{M}\, \Nbar \left[ \left( \Ximinus \right) , a_\nu
                  \right] \sigmamunu N
                  \nonumber\\[5pt]
& & \quad
    {}+ \frac{4 \beta_\pi}{M}\, \Nbar N \Tr \left[ \left( \Ximinus \right)
                  a^\mu \right]
    + i\, \frac{f_\rho g_\rho}{4 M} \, \Nbar \left[ \left( \Xiplus \right) ,
                  \rho_\nu \right] \sigmamunu N
                  \nonumber\\[5pt]
& & \quad
    {}+ 2i g_{\rho\pi\pi} \frac{\fpi^2}{m^2_\rho} \, \Tr \left\{ \rho^{\mu\nu}
                  \left[ \left( \Ximinus \right) , a_\nu \right]
             + v^{\mu\nu} \left[ \left( \Xiplus \right) , \rho_\nu \right]
                  \right\} \nonumber\\[3pt]
& & \quad
    {}+ i \Tr \left\{ \left[ \left( \Xiplus \right) ,
                  \rho_\nu \right] \rho^{\mu\nu} \right\} \ ,
                  \label{eq:Jmin}
\end{eqnarray}
\begin{eqnarray}
\lagrang^\mathrm{min}_{e^2} &\!\!=\!\!&
   e^2 A_\mu A^\mu \, \frac{\fpi^2}{4} \left( 1 + \frac{4 \beta_\pi}
          {\fpi^2 M} \, \Nbar N \right) \,
          \left[ 1 - \frac{1}{2}
                 \Tr \left( \tau_3 U \tau_3 U^\dagger \right) \right]
         \nonumber\\[5pt]
& & \quad
  {} + \frac{e^2}{4} \left( A_\lambda A^\lambda g^{\mu\nu} -
   A^\mu A^\nu \right) \Tr \left\{ \left[ \left( \Xiplus
               \right), \rho_\mu \right] \left[ \left( \Xiplus \right) ,
               \rho_\nu \right] \right\}  \nonumber\\[5pt]
& & \quad
  {}+ e^2 g_{\rho\pi\pi} \frac{\fpi^2}{m^2_\rho} \left( A_\lambda
              A^\lambda g^{\mu\nu} - A^\mu A^\nu \right)
              \nonumber\\[3pt]
& & \qquad\qquad\qquad\qquad \times
              \Tr \left\{ \left[ \left( \Xiplus \right) , \rho_\mu \right]
              \left[ \left( \Ximinus \right) , a_\nu \right]
              \right\} \ . \label{eq:Lesqmin}
\end{eqnarray}
The EM charge operator can be computed using methods analogous to
those in Ref.~\cite{AXC}, and it depends only on products of the
fields and their conjugate momenta \cite{Wei95}:
\begin{equation}
Q = \int\dthreex \left[ \frac{1}{2}\, N^\dagger (1 + \tau_3 ) N
          + ( \bm{\pi} \, \bm{\times} \, \bm{P}_\pi )_3
          + ( \bm{\rho}_\nu \, \bm{\times} \, \bm{P}_\rho^\nu )_3 \right]
          \ , \label{eq:EMcharge}
\end{equation}
where $P^a_\pi \equiv \partial {\cal L}/ \partial (\partial_0
\pi^a)$ and $(P_\rho)^{a\mu} \equiv \partial {\cal L}/ \partial
(\partial_0 \rho^a_\mu )$.

There are several relevant observations to be made about the results
in Eqs.~(\ref{eq:Jmin}) and (\ref{eq:Lesqmin}).  Evidently,
$J^\mu_\mathrm{min} = \frac{1}{2} B^\mu + V^\mu_3$, with $B^\mu$
from Eq.~(\ref{eq:baryoncurr}) and $\mathbf{V}^\mu$ from
Eq.~(\ref{eq:Vtotal}); thus, $J^\mu_\mathrm{min}$ is conserved
through $O(e^0)$. The results written in
Eqs.~(\ref{eq:eft-lagrangian}), (\ref{eq:LEMmin}), (\ref{eq:Jmin}),
and (\ref{eq:Lesqmin}) are the most efficient for generating the
Feynman rules, since they represent an explicit expansion in powers
of the electric charge $e$.  Nevertheless, the individual parts of
the lagrangian are \emph{not} EM gauge invariant by themselves; only
$\widetilde{{\cal L}}_{\mathrm{EFT}} \equiv {\cal L}_{\mathrm{EFT}}
+ \lagrang^\mathrm{min}_\mathrm{EM}$ is. In particular, the $O(e^2)$
``seagull'' terms involving mesons and two photons are crucial for
maintaining gauge invariance.  Most importantly, although the
minimal current of Eq.~(\ref{eq:Jmin}) is conserved through
$O(e^0)$, it is not exactly conserved due to the EM interactions:
$\partial_\mu J^\mu_\mathrm{min} =
\partial_\mu V^\mu_3 = O(e) \not= 0$. (The baryon current $B^\mu$ remains
conserved.) Thus we \emph{cannot} identify $J^\mu_\mathrm{min}$ as
the EM current.

To find the conserved, minimal current, we must use Noether's
theorem on the gauged lagrangian $\widetilde{{\cal
L}}_{\mathrm{EFT}}$.  A moment of reflection will convince the
reader that the desired result can be obtained by simply replacing
the ordinary derivatives in $J^\mu_\mathrm{min}$ with
gauge-covariant derivatives:
\begin{eqnarray}
{\widetilde J}^\mu_\mathrm{min} &\!\!=\!\!&
    -i\, \frac{\fpi^2}{4} \Tr [\tau_3 (U \covdev^{\mkern2mu\mu}
    U^\dagger + U^\dagger
                             \covdev^{\mkern2mu\mu} U )]
    + \frac{1}{2}\, \Nbar \gammamu \left( 1 + \Xiplus \right) N
               \nonumber\\[5pt]
& & \quad
    {}+ \frac{1}{2}\, g_A \Nbar \gammamu\gammafivel \left( \Ximinus \right) N
    + i\, \frac{\kappa_\pi}{M}\, \Nbar \left[ \left( \Ximinus \right) , a_\nu
                  \right] \sigmamunu N
                  \nonumber\\[5pt]
& & \quad
    {}+ \frac{4 \beta_\pi}{M}\, \Nbar N \Tr \left[ \left( \Ximinus \right)
                  {\widetilde a}^{\mkern2mu\mu} \right]
    + i\, \frac{f_\rho g_\rho}{4 M} \, \Nbar \left[ \left( \Xiplus \right) ,
                  \rho_\nu \right] \sigmamunu N
                  \nonumber\\[5pt]
& & \quad
    {}+ 2i g_{\rho\pi\pi} \frac{\fpi^2}{m^2_\rho} \, \Tr \left\{
    {\widetilde \rho}^{\mkern2mu\mu\nu}
                  \left[ \left( \Ximinus \right) , a_\nu \right]
             + {\widetilde v}^{\mkern2mu\mu\nu}
             \left[ \left( \Xiplus \right) , \rho_\nu \right]
                  \right\}  \nonumber\\[5pt]
& & \quad
    {}+ i \Tr \left\{ \left[ \left( \Xiplus \right) ,
                  \rho_\nu \right] {\widetilde \rho}^{\mkern2mu\mu\nu}
                  \right\} \nonumber\\[5pt]
&\!\!=\!\!& \frac{1}{2} B^\mu + {\widetilde V}^\mu_3 \ .
\label{eq:Jmintwiddle}
\end{eqnarray}
There are four important things to note about this result.  First,
the leading-order nucleon terms are the same as in
$J^\mu_\mathrm{min}$, since they contain no derivatives.  Second,
all factors of $\tau_3$ appear in the combinations $\xi \tau_3
\xi^\dagger$ or $\xi^\dagger \tau_3 \xi$ (except when they are
combined with $U$).  Third, there are no explicit factors of
${\widetilde v}_\mu$; these are all hidden inside the field tensors
${\widetilde \rho}^{\mkern2mu\mu\nu}$ and ${\widetilde
v}^{\mkern2mu\mu\nu}$.  Finally, there is no need to use
${\widetilde a}_\mu$ when it is inside a commutator; the $O(e)$
corrections vanish identically by Eq.~(\ref{eq:atwiddle}).

To prove that ${\widetilde J}^\mu_\mathrm{min}$ is indeed the
unique, conserved, minimal EM current, it suffices to evaluate the
Euler--Lagrange equations for the photon field based on
$\widetilde{{\cal L}}_{\mathrm{EFT}}$.  One finds
\begin{equation}
\partial_\nu F^{\nu\mu} =  e \left( J^\mu_\mathrm{min} -
    \frac{1}{e}\, \frac{\partial \lagrang^\mathrm{min}_{e^2}}{\partial
    A_\mu} \right)
    = e \,{\widetilde J}^\mu_\mathrm{min} \ ,
   \label{eq:Maxwell}
\end{equation}
where the final equality follows from Eq.~(\ref{eq:Jmintwiddle}) and
straightforward algebraic manipulation of Eqs.~(\ref{eq:Jmin}) and
(\ref{eq:Lesqmin}).  So we observe immediately that ${\widetilde
J}^\mu_\mathrm{min}$ is in fact the source term in Maxwell's
equations, and that
\begin{equation}
\partial_\mu {\widetilde J}^\mu_\mathrm{min} = 0 \ ,
\end{equation}
consistent with its identification as the EM current.  Note that
this current is conserved only for fields that satisfy the
Euler--Lagrange equations.  Moreover, by adding and subtracting
terms containing $(\partial \lagrang^\mathrm{min}_{e^2} / \partial
A_\mu)$, we can rewrite Eq.~(\ref{eq:LEMmin}) as
\begin{equation}
\lagrang^\mathrm{min}_\mathrm{EM} =
        -\frac{1}{4} F_{\mu\nu} F^{\mu\nu} - e A_\mu
        {\widetilde J}^\mu_\mathrm{min} - A_\mu \frac{\partial
        \lagrang^\mathrm{min}_{e^2}}{\partial A_\mu}
          + \lagrang^\mathrm{min}_{e^2} \ ,
\label{eq:LEMminnew}
\end{equation}
although the utility of this result is not immediately obvious.

What has become of the isovector currents $\mathbf{V}^\mu$ and
$\mathbf{A}^\mu$? It is instructive to begin by studying the
residual symmetries of $\widetilde{{\cal L}}_{\mathrm{EFT}}$. It is
clear that with the addition of the EM interactions, the gauged
currents ${\widetilde{\mathbf{V}}}^\mu$ and
${\widetilde{\mathbf{A}}}^\mu$ are no longer isovectors.
Nevertheless, as discussed in the Introduction, massless, two-flavor
QCD with EM interactions still possesses a residual, global, chiral
symmetry $U(1)_{L_3} \times U(1)_{R_3}$, where the left- and
right-handed rotations are around the third axis in isospin space.
We now show that this is indeed a symmetry of the gauged lagrangian
$\widetilde{{\cal L}}_{\mathrm{EFT}}$.

For this purpose, it is convenient to consider the lagrangian in the
form of Eqs.~(\ref{eq:eft-lagrangian}), (\ref{eq:LEMmin}),
(\ref{eq:Jmin}), and (\ref{eq:Lesqmin}).  The original lagrangian
${\cal L}_{\mathrm{EFT}}$ is invariant under the full chiral group
$SU(2)_L \times SU(2)_R$, so it remains invariant under the residual
symmetry.  Moreover, the terms in
$\lagrang^\mathrm{min}_\mathrm{EM}$ involving factors of $U$ and
$U^\dagger$ are clearly invariant, since these factors transform
globally with the matrices $L_3$ and $R_3$, which commute with
$\tau_3$.  Finally, for left- and right-handed rotations restricted
to the third axis in isospin space, we can reduce the original field
transformations \cite{FST97,LNP641} to
\begin{eqnarray}
\xi (x) \to \xi' (x) &\! = \!& L_3 \xi (x) {\widetilde
    h}^\dagger (x) =
    {\widetilde h} (x) \xi (x) R^\dagger_3 \ , \label{eq:xitransf}
    \\[5pt]
N(x) \to N' (x) &\! = \!& {\widetilde h} (x) N (x) \ ,
    \label{eq:Ntransf}
    \\[5pt]
\rho_\mu (x) \to \rho'_\mu (x) &\! = \!& {\widetilde h} (x)
    \rho_\mu (x) {\widetilde h}^\dagger (x) \ . \label{eq:rhotransf}
\end{eqnarray}
(The photon field is unchanged, as are the isoscalar sigma and omega
fields.)  Here we use ${\widetilde h}(x)$ to denote local $SU(2)_V$
transformations in the restricted case; note that even though only
global rotations $L_3$ and $R_3$ are considered, the matrix
${\widetilde h}(x)$ will generally involve isospin rotations in
other directions.

We can now make the following observations.  First, the nucleon
field transforms as in Eq.~(\ref{eq:Ntransf}); there are no
derivatives of the nucleon field and no explicit factors of
$v_\mu$ in $\lagrang^\mathrm{min}_\mathrm{EM}$. Second, based on
Eqs.~(\ref{eq:xitransf}) and (\ref{eq:rhotransf}), all the remaining
meson tensors: $a_\mu$, $\rho_\mu$, $\rho_{\mu\nu}$, and
$v_{\mu\nu}$ transform \emph{homogeneously}. For example,
\begin{equation}
\rho_{\mu\nu} \to \rho'_{\mu\nu} = {\widetilde h} \rho_{\mu\nu}
   {\widetilde h}^\dagger \ .  \label{eq:rhomunutransf}
\end{equation}
All that remains is to examine the pion field combinations
\begin{equation}
{\cal Q}_\pm \equiv
  \xi \frac{\tau_3}{2} \xi^\dagger \pm \xi^\dagger
  \frac{\tau_3}{2} \xi \ .
  \label{eq:buffered}
\end{equation}
These are the only combinations of pion fields that have no
derivatives, are hermitian, conserve parity, and maintain the
residual symmetry. (The parity of ${\cal Q}_\pm$ is $\pm$.) The
proof of the homogeneous transformation property is simple:
\begin{eqnarray}
\xi^\dagger \tau_3 \xi &\to & {\xi'}^\dagger \tau_3 \xi' =
  ({\widetilde h} \xi^\dagger L^\dagger_3 ) \tau_3 (L_3 \xi
  {\widetilde h}^\dagger ) = {\widetilde h} ( \xi^\dagger \tau_3 \xi
  ) {\widetilde h}^\dagger \ , \label{eq:buffone} \\[5pt]
\xi \tau_3 \xi^\dagger &\to & \xi' \tau_3 {\xi'}^\dagger =
  ({\widetilde h} \xi R^\dagger_3 ) \tau_3 ( R_3 \xi^\dagger
  {\widetilde h}^\dagger ) = {\widetilde h} ( \xi \tau_3 \xi^\dagger
  ) {\widetilde h}^\dagger \ . \label{eq:bufftwo}
\end{eqnarray}
Note that this proof does not utilize the form of ${\widetilde h}$
and relies only on $[ L_3 , \tau_3 ] = 0 = [ R_3 , \tau_3 ]$.

The proof of the residual invariance of $\widetilde{{\cal
L}}_{\mathrm{EFT}}$ is now immediate and follows by inspection of
Eqs.~(\ref{eq:Jmin}) and (\ref{eq:Lesqmin}); $J^\mu_\mathrm{min}$
and $\lagrang^\mathrm{min}_{e^2}$ are independently invariant.  The
gauged lagrangian $\widetilde{{\cal L}}_{\mathrm{EFT}}$ admits three
conserved currents, one of which is $B^\mu$.  The other two
conserved currents are the gauged neutral currents ${\widetilde
V}^\mu_3$ and ${\widetilde A}^\mu_3$; the corresponding charged
currents ${\widetilde V}^\mu_\pm$ and ${\widetilde A}^\mu_\pm$ are
not conserved. In fact, the explicit result follows from a theorem
proven long ago by Adler and Coleman \cite{Adler65}, which in our
case (and with our notation) reads\footnote{%
An alternative way to write these relations is $\covdev_\mu
{\widetilde V}^\mu_a = 0 = \covdev_\mu {\widetilde A}^\mu_a$.}
\begin{equation}
\partial_\mu {\widetilde V}^\mu_a = \epsilon_{a3b}\, e A_\mu
{\widetilde V}^\mu_b \ , \qquad
\partial_\mu {\widetilde A}^\mu_a = \epsilon_{a3b}\, e A_\mu
{\widetilde A}^\mu_b \ . \label{eq:AdlerColeman}
\end{equation}
Note that the theorem does not assume that the field transformations
are linear, and it is valid only for fields that satisfy the
Euler--Lagrange equations. The divergence of the axial-vector
current ${\widetilde{\mathbf{A}}}^\mu$ omits contributions from
chiral anomalies, as well as from the explicit breaking of chiral
symmetry.  If the latter were included, we would have the PCAC
relation
\begin{equation}
\partial_\mu {\widetilde A}^\mu_a \propto m^2_\pi \pi_a + O(e) \ .
\label{eq:PCAC}
\end{equation}

Since the charged currents are no longer conserved, the
corresponding charges are time dependent.  The only constants of the
motion are the neutral charges, and the chiral charge algebra
reduces to the mutually commuting charges
\begin{equation}
[B, Q_3 ] = [B, (Q_5)_3 ] = [Q_3 , (Q_5)_3 ] = [Q, (Q_5)_3 ] = 0.
\label{eq:chargealgebra}
\end{equation}

To end this section, we discuss the relationship between our
procedure and the one that uses external fields \cite{GL84}. In the
external-field procedure, one includes EM interactions by
introducing spurious charge operators that transform under the full
chiral symmetry as
\begin{equation}
Q_L \to Q'_L = L Q_L L^\dagger \ , \qquad Q_R \to Q'_R = R Q_R
    R^\dagger \ . \label{eq:Qspurious}
\end{equation}
This permits the construction of so-called ``spurion'' fields
\cite{EGPdR,NR96} that transform homogeneously under the full
symmetry group:
\begin{equation}
{\cal Q}_L \equiv \xi^\dagger Q_L \xi \ , \qquad {\cal Q}_R \equiv
\xi Q_R \xi^\dagger \ , \qquad {\cal Q}_{L,R} \to {\cal Q}'_{L,R} =
h {\cal Q}_{L,R} h^\dagger \ . \label{eq:spuriousfields}
\end{equation}
[Compare Eqs.~(\ref{eq:buffone}) and (\ref{eq:bufftwo}).] One then
constructs the most general (non-redundant) lagrangian to a given
order in derivatives, and at the end, replaces the spurious charge
operators $Q_{L,R}$ with the true charge operator $Q$ to produce a
result with the appropriate residual symmetries. Because of the
simplicity of the residual symmetry in the present case, it is clear
that the appropriate pion field operators (with well-defined parity)
are those in Eq.~(\ref{eq:buffered}).  Thus our procedure for
constructing $\lagrang_\mathrm{EM}$ is equivalent to the procedure
using external fields.

As a final note, since we have shown earlier that
$\lagrang^\mathrm{min}_{e^2}$ is independently invariant under the
residual chiral symmetry, when one computes $O(e^2)$ contributions
with photon loops from this lagrangian, one will require additional,
\emph{non-minimal} counterterms to render the calculations finite.
For instance, if one considers the purely pionic term in
Eq.~(\ref{eq:Lesqmin}), the leading-order (in derivatives)
counterterm can be deduced by ``integrating out'' the photon fields,
producing (up to an irrelevant additive constant)
\begin{equation}
{\cal L}^{(0)}_{e^2} = e^2 \, C \Tr (\tau_3 U \tau_3 U^\dagger ) \ .
\end{equation}
This single counterterm reproduces the well-known $SU(2)$ result
\cite{EGPdR,Scherer03}.

\section{Non-Minimal Couplings}
\label{sec:nonminimal}

Here we discuss couplings of the $\pi$ and $N$ to the photon that
are non-minimal and involve the field tensor $F_{\mu\nu}$ and its
derivatives.  These terms will be individually EM gauge invariant
and are relevant for describing the EM structure of nucleons and
pions order-by-order in a derivative expansion \cite{FST97}.  We
will consider non-minimal terms only for the hadronically stable
particles and return to discuss VMD contributions in the next
section.

We begin by considering a generalization of the non-minimal baryon
lagrangian proposed by Rusnak and Furnstahl \cite{RF97}:
\begin{eqnarray}
{\cal L}^\mathrm{had}_\mathrm{RF} &\! = \!& {}-\frac{e}{4M}\,
    F_{\mu\nu} \Nbar \lambda \sigma^{\mu\nu} N - \frac{e}{2M^2} \,
    (\partial^\nu  F_{\mu\nu}) \Nbar \beta \gammamu N \nonumber
    \\[5pt]
& & \quad
   {} - \frac{e}{4 M^3}\, (\partial_\nu \partial^\eta
   F_{\mu\eta}) \Nbar \lambda' \sigma^{\mu\nu} N - \frac{e}{M^4}\,
   (\partial^2 \partial^\nu F_{\mu\nu}) \Nbar \beta' \gammamu N +
   \cdots \ , \label{eq:RFlagrang}
\end{eqnarray}
where
\begin{equation}
\lambda = \lambda_\mathrm{p} \frac{1}{2} \, (1 + \tau_3 ) +
    \lambda_\mathrm{n} \frac{1}{2} \, (1 - \tau_3) = \frac{1}{2}
    \,(\lambda_\mathrm{p} + \lambda_\mathrm{n} ) + \frac{1}{2}
    \,(\lambda_\mathrm{p} - \lambda_\mathrm{n} ) \tau_3 \equiv
    \lambda^{(0)} + \lambda^{(1)}\tau_3 \ ,
\end{equation}
and similarly for $\beta$, $\lambda'$, and $\beta'$.  Here the
superscripts in parentheses denote the isospin, and
$\lambda_\mathrm{p} = 1.793$ and $\lambda_\mathrm{n} = -1.913$. The
constant $\lambda'$ contributes to $r_\mathrm{rms}$ of the anomalous
$(F_2)$ EM form factor, and $\beta'$ contributes to the $Q^4$
dependence of the charge $(F_1)$ EM form factor.  These two terms
involve $\nu > 4$ and will be redundant when we include the VMD part
of the lagrangian, so we will not consider them further in the
sequel.

While Eq.~(\ref{eq:RFlagrang}) is clearly EM gauge invariant (and
baryon phase invariant), it does not obey the residual $U(1)_{L_3}
\times U(1)_{R_3}$ chiral symmetry of two-flavor, massless QCD.  To
maintain this symmetry, we take instead
\begin{equation}
{\cal L}^\mathrm{had}_\mathrm{EM(N)} =  {}-\frac{e}{4M}\,
    F_{\mu\nu} \Nbar\, {\widetilde \lambda} \sigma^{\mu\nu} N
    - \frac{e}{2M^2} \,
    (\partial^\nu  F_{\mu\nu}) \Nbar\, {\widetilde \beta} \gammamu N
    \ , \label{eq:LhadN}
\end{equation}
where
\begin{equation}
{\widetilde \lambda} \equiv
    \lambda^{(0)} + \lambda^{(1)} \left( \Xiplus \right)
    = \lambda^{(0)} + \lambda^{(1)} {\cal Q}_+ \ ,
    \label{eq:lambdadef}
\end{equation}
and similarly for ${\widetilde \beta}$.  The factors ${\widetilde
\lambda}$ and ${\widetilde \beta}$ include the appropriate
positive-parity combination of pion fields, reduce to the
conventional constants when $\bm{\pi} \to 0$, and contain the fields
that would arise naturally using the external-field construction
discussed in the previous section.  Evidently, since $[q, \tau_3 ] =
0$, Eq.~(\ref{eq:LhadN}) remains EM gauge invariant as well, even
with ordinary derivatives.  Partial differentiation and evaluation
of $(\partial {\cal L}^\mathrm{had}_\mathrm{EM(N)}/
\partial A_\mu)$ allows one to identify the contribution to the EM
current (in agreement with the source term in Maxwell's equations),
which we illustrate below.

We now consider non-minimal couplings between pions and the EM
field. These were not discussed explicitly in Ref.~\cite{FST97},
since VMD implies that a coupling to rho mesons describes the pion
EM form factor quite accurately.  We return to this point in the
next section.

The lowest-order $(\nu = 4)$, non-minimal pion--photon couplings are
\begin{equation}
{\cal L}^\mathrm{had}_{\mathrm{EM}(\pi )} =  e \,\omega_1 \Tr
   \left[ \left( \Xiplus \right) {\widetilde v}^{\mkern2mu\mu\nu}
   \right] \!F_{\mu\nu} + e \,\omega_2 \Tr \left[ \left( \Ximinus
   \right) {\widetilde a}^{\mkern2mu\mu} \right] \!\partial^\nu
   F_{\mu\nu} \ . \label{eq:Lhadpi}
\end{equation}
%
This lagrangian is clearly invariant under the local $U(1)_Q$ charge
symmetry. Moreover, since ${\widetilde v}^{\mkern2mu\mu\nu}$ and
${\widetilde a}^{\mkern2mu\mu}$ transform homogeneously under the
residual chiral symmetry [see Eqs.~(\ref{eq:atwiddle}) and
(\ref{eq:covvmunu})], this is also respected.

Partial integration again allows for a determination of the pionic
contribution to the non-minimal EM current. When combined with the
non-minimal nucleon current discussed above, we find
\begin{equation}
{\cal L}^\mathrm{had}_\mathrm{EM} =  -e A_\mu {\widetilde
J}^\mu_\mathrm{had} \ , \label{eq:Lemhaddef}
\end{equation}
\begin{eqnarray}
{\widetilde J}^\mu_\mathrm{had} &\! = \!& \frac{1}{2M}\,
    \partial_\nu ( \Nbar\, {\widetilde \lambda} \sigma^{\mu\nu} N) -
    \frac{1}{2M^2}\, \left( g^{\mu\nu} \partial^2 -
    \partial^\mu \partial^\nu \right) ( \Nbar \,
    {\widetilde \beta} \gamma_\nu N) \nonumber \\[6pt]
%
& & \quad {}- 2 \omega_1 \,\partial_\nu \Tr \left[ \left( \Xiplus
   \right) {\widetilde v}^{\mkern2mu\mu\nu} \right]
   \nonumber \\[6pt]
& & \quad {}+ \omega_2  \left( g^{\mu\nu} \partial^2 -
    \partial^\mu \partial^\nu \right) \Tr
   \left[ \left( \Ximinus
   \right) {\widetilde a}_\nu
   \right]  \ . \label{eq:Jhaddef}
\end{eqnarray}
Note that $\partial_\mu {\widetilde J}^\mu_\mathrm{had} = 0$ follows
by inspection.  Moreover, since this relation holds
\emph{identically}, it is true whether or not the fields satisfy the
Euler--Lagrange equations. Since the baryon current of
Eq.~(\ref{eq:baryoncurr}) remains conserved,
\begin{equation}
\partial^\nu ( \Nbar \,
    {\widetilde \beta} \gamma_\nu N) = \beta^{(1)} \partial^\nu
    \left[ \Nbar \left( \Xiplus \right) \gamma_\nu N \right] \ .
\end{equation}

The lagrangian of Eq.~(\ref{eq:LhadN}) contains no derivatives on
the baryon fields, so it gives no new contributions to $B^\mu$,
${\widetilde{\mathbf{V}}}^\mu$, or ${\widetilde{\mathbf{A}}}^\mu$.
The EM current determined thus far is given by\footnote{%
Although the lagrangian of Eq.~(\protect\ref{eq:Lhadpi}) generates
additional contributions to ${\widetilde{\mathbf{V}}}^\mu$ and
${\widetilde{\mathbf{A}}}^\mu$, these are unrelated to the pionic
part of ${\widetilde J}^\mu_\mathrm{had}$, are not identically
conserved, and are explicitly of $O(e)$.  Thus they are not of
particular interest.}
\begin{equation}
{\widetilde J}^\mu_\mathrm{min} + {\widetilde J}^\mu_\mathrm{had} =
    \frac{1}{2} B^\mu + {\widetilde V}^\mu_3 + {\widetilde
    J}^\mu_\mathrm{had} \ . \label{eq:Jboth}
\end{equation}
Note that ${\widetilde J}^\mu_\mathrm{min}$ and ${\widetilde
J}^\mu_\mathrm{had}$ are \emph{independently} conserved, the latter
identically and the former by virtue of the Euler--Lagrange
equations.  Moreover, since the non-minimal terms respect the
residual, global chiral symmetry, $\partial_\mu {\widetilde A}^\mu_3
= 0$ remains true as well. The freedom to add ${\widetilde
J}^\mu_\mathrm{had}$, which generally contains both isoscalar and
isovector parts, reflects the \emph{non-uniqueness} of the EM
current in the effective theory.  One can always augment the unique
minimal current ${\widetilde J}^\mu_\mathrm{min}$ by terms that are
independently conserved, without spoiling the conservation of the
total EM current.  Thus the relationship between the electromagnetic
current and $\frac{1}{2}B^\mu + V^\mu_3$, valid for $u$ and $d$
quarks, is modified in the effective field theory. Note, however,
that $Q$ is still given by Eqs.~(\ref{eq:Qdef}) and
(\ref{eq:EMcharge}); since ${\widetilde J}^\mu_\mathrm{had}$ is
conserved \emph{identically}, it does not produce a new symmetry
generator.

\section{Vector Meson Dominance}
\label{sec:VMD}

There are two basic assumptions underlying the formalism of vector
meson dominance:
\begin{itemize}
\item
Photon interactions with hadrons are mediated primarily by the
exchange of low-mass, neutral vector mesons \cite{SAKURAI60,KLZ}.
\item
One can describe processes involving \emph{spacelike} photons and
vector mesons using photon--meson (and meson--meson) couplings
determined from hadronic decay widths, in which the meson
four-momentum is \emph{timelike}.  This hypothesis was justified
using dispersion relations in Ref.~\cite{GMZ}.
\end{itemize}

For the rho meson, we can start with the expression in
Ref.~\cite{KLZ}:
\begin{equation}
{\cal L}_\rho = -\frac{e}{2 g_\gamma}\, (\partial^\mu \rho^\nu_3 -
\partial^\nu \rho^\mu_3 ) F_{\mu\nu} \ , \label{eq:KLZrho}
\end{equation}
where $g_\gamma$ (denoted by $g_\rho$ in \cite{KLZ}) is determined
from $\rho^0 \to e^+ e^-$ decay.  This expression can be extended to
fully reflect the EM gauge invariance and residual chiral symmetry
by using
\begin{equation}
{\cal L}^\mathrm{vmd}_\rho = -\frac{e}{2 g_\gamma} \Tr \left[
   \left( \Xiplus \right) \rho^{\mu\nu} \right] F_{\mu\nu} \ .
   \label{eq:vmdrho}
\end{equation}
Note that we can use $\rho^{\mu\nu}$ here rather than ${\widetilde
\rho}^{\mkern2mu\mu\nu}$, since the $O(e)$ corrections produced by
the latter vanish identically [see Eq.~(\ref{eq:covDrho})]. This
implies that this VMD term involves only a direct coupling between
the photon and rho (and pions), \emph{without any seagulls}
involving two photons.

The expression for the $\rho^0 \to e^+ e^-$ decay width is
\cite{SAKURAI69} ($\alpha$ is the fine-structure constant)
\begin{equation}
\Gamma (\rho^0 \to e^+ e^- ) = \frac{\alpha^2}{g^2_\gamma /4\pi}
   \left( \frac{m_\rho}{3} \right) \left[ 1 + 2 \left(
   \frac{m_e}{m_\rho} \right)^2 \right] \left[ 1 - 4 \left(
   \frac{m_e}{m_\rho} \right)^2 \right]^{1/2} \ .
   \label{eq:rhodecay}
\end{equation}
Using the experimental values \cite{PDG}\ $\Gamma (\rho^0 \to e^+
e^- ) = 7.02 \pm 0.11 \, \mathrm{keV}$ and $m_\rho = 776 \,
\mathrm{MeV}$, we find $g^2_\gamma /4\pi = 1.96$, which agrees with
Ref.~\cite{FST97}.

For the omega meson, we again start with the analysis of
Ref.~\cite{KLZ} and write
\begin{equation}
{\cal L}_0 = \frac{e}{2}\, \frac{1}{2 g_{\sssize Y}} \left( \cos
   \theta_Y
   \Phi^{\mu\nu} - \sin \theta_Y V^{\mu\nu} \right) F_{\mu\nu}
   \ , \label{eq:KLZhyper}
\end{equation}
where $\Phi^{\mu\nu} = \partial^\mu \Phi^\nu - \partial^\nu
\Phi^\mu$ and $V^{\mu\nu} = \partial^\mu V^\nu -
\partial^\nu V^\mu$
are the field tensors for the $\phi(1020)$ and $\omega(782)$,
respectively.  We will not consider the $\phi(1020)$ coupling
further, since it is ``integrated out'' in our EFT lagrangian, as we
discuss below.

Thus we are considering only
\begin{equation}
{\cal L}_\omega = -\frac{e}{2} \left( \frac{\sin \theta_Y}
   {2 g_{\sssize Y}} \right) V^{\mu\nu} F_{\mu\nu} \ .
   \label{eq:KLZomega}
\end{equation}
As discussed by Sakurai \cite{SAKURAI69}, exact $SU(3)_f$ symmetry
(denoted by the accent $\circ$) implies that
${\skew3\mathring\theta}_Y = 0$, the ${\skew4\mathring\phi}$ is pure
octet, the ${\skew1\mathring\omega}$ is pure singlet, and the
hypercharge and isospin couplings are related by
\begin{equation}
{\skew2\mathring{g}}_{\sssize Y} = \frac{\sqrt{3}}{2}\,
{\skew2\mathring{g}}_\gamma \ . \label{eq:gcirc}
\end{equation}
This relation follows from the $SU(3)$ Clebsch--Gordan coefficients
(for pure $F$-type photon--meson couplings).  Thus, to have
phi--omega mixing, \emph{there must be explicit} $SU(3)_f$
\emph{symmetry breaking.}

If we assume that the explicit symmetry breaking is described by
``ideal mixing'', so that the physical $\phi(1020)$ contains only
strange (valence) quarks, we find $\sin \theta_Y = 1/ \sqrt{3}$.
Since this mixing is due to the explicit $SU(3)_f$ symmetry
breaking, we can neglect higher-order symmetry-breaking effects by
using Eq.~(\ref{eq:gcirc}) to evaluate $g_{\sssize Y}$, which
produces
\begin{equation}
{\cal L}_\omega = -\frac{e}{2 g_\gamma} \left( \frac{1}
   {3} \right) V^{\mu\nu} F_{\mu\nu} \ ,
   \label{eq:idealomega}
\end{equation}
the result used in Ref.~\cite{FST97}.

One can measure $\theta_Y$ by comparing the decays $\omega \to e^+
e^-$ and $\phi(1020) \to e^+ e^-$ \cite{DSharp}.  Since the leptonic
decays are given by Eq.~(\ref{eq:rhodecay}), with appropriate
substitution of masses and couplings, one finds
\begin{equation}
\frac{\Gamma (\omega \to e^+ e^- )}{\Gamma (\phi \to e^+ e^- )} =
\frac{m_\omega}{m_\phi} \, \tan^2 \theta_Y \ . \label{eq:DSharp}
\end{equation}
Using the empirical values \cite{PDG}\ $\Gamma (\omega \to e^+ e^- )
= 0.60 \pm 0.02 \, \mathrm{keV}$, $\Gamma (\phi \to e^+ e^- ) = 1.27
\pm 0.02 \, \mathrm{keV}$, $m_\omega = 783 \, \mathrm{MeV}$, and
$m_\phi = 1019 \, \mathrm{MeV}$, we find $(\sin \theta_Y
)_\mathrm{expt} = 0.617 \pm 0.009$, which differs from the
ideal-mixing result $1/\sqrt{3}$ by only 7\%.  So the real world is
close to ideal mixing, and Eq.~(\ref{eq:idealomega}) is adequate for
parametrizing hadronic EM form factors.

Thus we take as our VMD lagrangian the sum of Eqs.~(\ref{eq:vmdrho})
and (\ref{eq:idealomega}):
\begin{equation}
{\cal L}^\mathrm{vmd}_\mathrm{EM}  = -\frac{e}{2 g_\gamma}
   \left\{\Tr \left[
   \left( \Xiplus \right) \rho^{\mu\nu} \right]  + \frac{1}{3} \,
   V^{\mu\nu}
   \right\} F_{\mu\nu} \ . \label{eq:LEMvmd}
\end{equation}
[Compare Eq.~(52) in Ref.~\cite{FST97}.] This lagrangian is
invariant under the local EM gauge symmetry and the residual, global
$U(1)$ chiral symmetry.  The factor (1/3) for the $\omega\gamma$
coupling is valid if we assume ideal mixing and keep only the
leading-order explicit $SU(3)_f$ symmetry-breaking effects. We note
in passing that enforcing the residual chiral symmetry leads to a
$\rho\pi\pi\gamma$ coupling that can contribute to a two-nucleon,
$\rho\pi$ exchange current in pion photoproduction, in which all the
couplings are known from other processes.  Vector meson dominance
and non-minimal EM couplings were used in Ref.~\cite{FST97} to
describe the contributions of nucleon EM structure to the
\emph{nuclear} charge form factors, without introducing \emph{ad
hoc} form factors at the NN$\gamma$ vertex.\footnote{%
If the $\omega_i$ parameters of Eq.~(\protect\ref{eq:Lhadpi}) are
included in the pion charge form factor along with the VMD
contribution, a fit to the experimental rms charge radius $0.66 \pm
0.01\,\mathrm{fm}$ \protect\cite{PIRAD1,PIRAD2} confirms that these
parameters are small: $\omega_1 + \omega_2 = (1.2 \pm 0.7) \times
10^{-3}$.} [See Eqs.~(56) through (68) in Ref.~\cite{FST97}.]

One might argue that in a theory restricted to the light-quark
$(u,d)$ sector, arguments based on $SU(3)_f$ symmetry are not
particularly relevant.  Indeed, we could simply introduce an
independent coupling $g_0$ and take
\begin{equation}
{\cal L}'_\omega = -\frac{e}{2 g_0}\, V^{\mu\nu} F_{\mu\nu} \ ,
   \label{eq:empomega}
\end{equation}
analogous to Eq.~(\ref{eq:KLZrho}).  We can determine $g_0$
empirically from $\omega \to e^+ e^-$ decay, or alternatively,
combine this with $\rho^0 \to e^+ e^-$ decay to find $g_0^2 = 11.8
\, g^2_\gamma$, or
\begin{equation}
g_0 = 3.44 \, g_\gamma \ . \label{eq:exptgs}
\end{equation}
So we could just use (1/3.44) in place of (1/3) in
Eq.~(\ref{eq:LEMvmd}) to reproduce $\Gamma (\omega \to e^+ e^- )$
precisely, but it is remarkable that the estimate based on ideal
mixing and lowest-order $SU(3)_f$ symmetry breaking gives such an
accurate result.  [Interestingly, if we simply set $g_{\sssize Y} =
g_\gamma$ in Eq.~(\ref{eq:KLZomega}), we would find $g^2_0 = 12
g^2_\gamma$, within 2\% of the experimental result!]

With the assumption of ideal mixing, the $\phi(1020)$ is composed
only of (valence) strange quarks.  Thus it has a weak coupling to
nucleons (also true empirically), and its mass is 30\% larger than
the masses of the $\rho$ and $\omega$. So it is entirely appropriate
to integrate out the $\phi(1020)$ degrees of freedom and omit them
from the VMD lagrangian (\ref{eq:LEMvmd}).  Within the explicit
$SU(3)_f$ breaking scenario described above, one then finds a
contribution to the $\beta^{(0)}$ parameter in ${\widetilde
J}^\mu_\mathrm{had}$ [Eq.~(\ref{eq:Jhaddef})] equal to $-\sqrt{2}M^2
g_\phi / 3 g_\gamma m^2_\phi$, where $m_\phi$ is the $\phi(1020)$
mass and $g_\phi$ is its coupling to nucleons.  Similarly, the
effects of the $\phi(1020)$ in the nucleon--nucleon interaction can
be absorbed in the (isoscalar) NN$\omega$ coupling parameter $g_v$
\cite{FST97}.

As a final comment, we note that it is possible to augment the VMD
couplings in Eq.~(\ref{eq:LEMvmd}) by multiplying the interactions
by isoscalar combinations like $\phi$, $\phi^2$, $V_\mu V^\mu$, etc.
(Here $\phi$ is the field of the $\sigma$.) These terms all have
$\nu \geq 5$ and contain at least two heavy bosons. Nevertheless,
they allow for the possibility of isoscalar EM exchange currents in
nuclei.

\section{Anomalies}
\label{sec:anomalies}

The lagrangian density $\lagrang^\mathrm{anom}_\mathrm{EM}$ contains
two types of terms involving abnormal-parity couplings: terms
arising from manifestly chirally invariant expressions (for pions,
these first appear at sixth order in derivatives\footnote{%
Here ``sixth order'' includes external fields, their derivatives,
and factors of $m^2_\pi$ in the counting.  This is equivalent to
$\nu = 6$ in our counting scheme.} \cite{EFS02,Scherer03}), and
terms arising from the Wess--Zumino--Witten action that describes
the chiral anomaly \cite{WessZumino71,Witten83,Wei96+}. The latter
can be expressed as a lagrangian density by using an infinite number
of terms that are not chirally invariant \cite{WessZumino71};
nevertheless, a chiral transformation produces a variation in these
terms that is a spacetime derivative, preserving the chiral
invariance of the action. All of these abnormal-parity terms contain
an odd number of pseudoscalar Goldstone bosons (perhaps coupled to
particles with normal parity) and are constrained by $G$-parity
\cite{Witten83,Wei96+}. Overall parity $(P)$ conservation is
restored by the presence of the antisymmetric tensor
$\epsilon^{\mu\nu\alpha\beta}$ in each term.

Anomalies arise because the fermion measure in the path integral for
QCD is not invariant under chiral transformations
\cite{Fuji79,Fuji80}. This implies that in a quantum field theory
with fermions coupled to vector or axial-vector fields, it is
generally impossible to satisfy \emph{simultaneously} the vector and
axial Ward identities derived from the lagrangian through Noether's
theorem. The anomalies evidently depend on the fermion couplings to
the vector and axial-vector fields, and the form of the anomalous
chiral action is well known \cite{Wei96+}.

Since the anomalies are perturbative, then just as in weak-coupling
theories, they can be computed exactly from the underlying QCD. The
structure and strength of the anomalies are determined by chiral
symmetry and the number of colors $N_c$ in QCD, together with a
specific regularization prescription. The results are general and
are not unique to the particular low-energy representation in
Eq.~(\ref{eq:eft-lagrangian}). The contributions from pseudoscalar
mesons alone and in combination with vector mesons are discussed,
for example, in Refs.~\cite{WessZumino71,Meissner88,Scherer03}. With
the power counting used in this work, anomalous interaction terms
start at order $\nu = 4$.

EM interactions can be included either by introducing external
fields \cite{WessZumino71} or by using Witten's ``trial and error''
method \cite{Witten83}.  There are terms of $O(e)$ and $O(e^2)$, as
required by EM gauge invariance.  They can be written using a
lagrangian density with a finite number of terms. All the
leading-order interaction terms involving pseudoscalar bosons and
photons are given by Wess and Zumino \cite{WessZumino71}; the
overall normalization is given in Eq.~(21) of Ref.~\cite{Witten83}.
Applications of the EM anomalous action have focused primarily on
meson decays.

For our purposes, the important point is that the anomalous EM terms
contain only bosons.  Thus they enter in electromagnetic
interactions with nuclei only through meson-exchange currents.
Moreover, the anomalous EM couplings either: (1) are of $O(e^2)$,
like $\pi^0 \to \gamma\gamma$; (2) contain (at least) three pions
$(\gamma^\ast \to \pi\pi\pi$, as in $\omega \to \gamma^\ast \to
\pi\pi\pi)$; or (3) involve a heavy boson $(\omega \to \pi^0
\gamma$, $\rho \to \pi \gamma)$.  The resulting abnormal-parity
exchange currents are not likely to be very important for studying
EM interactions in the nuclear many-body problem, although they may
be relevant in few-nucleon systems. (From an examination of the
experimental decay widths, the most important contribution probably
arises from the $\omega \pi^0 \gamma$ coupling.)

Based on these considerations, it is premature to enumerate
all of the $\nu = 4$ anomalous EM couplings in this
$SU(2) \times SU(2) \times U(1)$ QHD EFT. We relegate this task to
future work, when and if it is necessary.

\section{Summary}
\label{sec:summary}

In this work, we incorporate electromagnetic interactions in a
recently proposed hadronic lagrangian with a nonlinear realization
of chiral symmetry \cite{FST97}.  The effective lagrangian provides
a systematic framework for calculating both nuclear wave functions
and nuclear exchange currents.  The lagrangian is truncated by
working to a fixed order in the parameter $\nu$, which essentially
counts powers of ratios of the particle momenta to the nucleon mass
$M$ or of mean meson fields to the nucleon mass \cite{FST97,RF97}.
Practically speaking, in the nuclear many-body problem, the
expansion is in powers of $\kfermi / M$, where $\kfermi$ is the
Fermi wave number at equilibrium nuclear density; this ratio
provides a small parameter for ordinary nuclei and for electroweak
processes at modest momentum transfers.

For the degrees of freedom considered here (N, $\pi$, $\sigma$,
$\omega$, $\rho$), truncated at a specific order in terms of fields
and their derivatives $(\nu = 4)$, we construct the most general
(non-redundant) lagrangian consistent with Lorentz invariance; $P$,
$C$, and $T$ symmetry; electromagnetic gauge invariance; and the
residual chiral symmetry of two-flavor, massless QCD with
electromagnetic interactions. By introducing EM gauge-covariant
derivatives, we formulate the minimal lagrangian required by gauge
invariance.  This is given by Eqs.~(\ref{eq:LEMmin}),
(\ref{eq:Jmin}), and (\ref{eq:Lesqmin}), with the EM charge operator
of Eq.~(\ref{eq:EMcharge}). The conserved current at $O(e^0)$,
$J^\mu_\mathrm{min}$, is equal to the sum of the Noether currents
$\frac{1}{2} B^\mu + V^\mu_3$ of the original lagrangian ($V^\mu_3$
is given in the Appendix), but this current is no longer exactly
conserved due to the EM interactions.  The unique, conserved,
minimal EM current, ${\widetilde J}^\mu_\mathrm{min}$, is obtained
by replacing all ordinary derivatives in $V^\mu_3$ with
gauge-covariant derivatives; it is verified that ${\widetilde
J}^\mu_\mathrm{min}$ [Eq.~(\ref{eq:Jmintwiddle})] is indeed the
source in Maxwell's equations.

We show that the gauged neutral currents ${\widetilde V}^\mu_3$ and
${\widetilde A}^\mu_3$ are exactly conserved, but the gauged charged
currents ${\widetilde V}^\mu_\pm$ and ${\widetilde A}^\mu_\pm$ are
not; that is, $\covdev_\mu {\widetilde V}^\mu_a = 0 = \covdev_\mu
{\widetilde A}^\mu_a$. (Here $\covdev_\mu$ is the EM gauge-covariant
derivative.)  The conserved currents arise because of the residual,
global, chiral symmetry $U(1)_{L_3} \times U(1)_{R_3}$, where the
left- and right-handed rotations are around the third axis in
isospin space.  This symmetry obtains in part because the meson
charge matrix enters in the form ${\cal Q}_\pm$ of
Eq.~(\ref{eq:buffered}), which is identical to the form that would
arise in the external-field method of Ref.~\cite{GL84}.

Non-minimal couplings of the photon to the nucleon and pion are
introduced in Eqs.~(\ref{eq:LhadN}) and (\ref{eq:Lhadpi}). These
couplings are automatically gauge invariant because they depend on
the EM field tensor $F_{\mu\nu}$ and its derivatives, and they
respect the residual chiral symmetry because they contain the
factors ${\cal Q}_\pm$.  The enforcement of the residual chiral
symmetry implies additional interaction vertices (and meson-exchange
currents) containing pions. When combined with the
vector-meson-dominance couplings of Eq.~(\ref{eq:LEMvmd}), these
non-minimal couplings are known to adequately reproduce the pion and
nucleon EM form factors at low momentum transfers.  For the
isoscalar $\omega$ and $\phi (1020)$ mesons, we assume ideal mixing
and keep only the leading-order explicit $SU(3)_f$ symmetry-breaking
effects; this reproduces the experimental $\omega \to e^+ e^-$ decay
width to 30\%, which can be easily improved using
Eqs.~(\ref{eq:empomega}) and (\ref{eq:exptgs}).  The EM couplings
arising from the anomalous Wess--Zumino--Witten action are also
considered, but since these couplings are of order $\nu = 4$ and
contribute only in meson-exchange currents, their analysis is left
for a future project.

Other future projects based on this QHD EFT lagrangian will include:
(1) The computation of the Lorentz-covariant, one- and two-nucleon
amplitudes for electron scattering and pion photoproduction (and
electroproduction) that can be used in many-body calculations of
medium and heavy nuclei \cite{CEH}; (2) The inclusion of the $\Delta
(1232)$ as an explicit degree of freedom and the determination of
its EM interactions subject to the residual chiral symmetry
mentioned above; and (3) The extension of the non-minimal, hadronic
EM lagrangian to include higher-order terms in the derivative
expansion [as in Eq.~(\ref{eq:RFlagrang})], which will allow the
nucleon EM form factors to be accurately described at momentum
transfers large enough to study contributions from meson-exchange
currents in nuclei.

By working with the QHD EFT lagrangian, which uses the same degrees
of freedom to describe the one- and two-body currents and the
nuclear many-body dynamics, which respects the underlying symmetries
of QCD (both before and after EM interactions are included), and
which has parameters that can be calibrated from strong-interaction
phenomena (and meson decays), we have a consistent, self-contained,
Lorentz-covariant field-theory framework in which to carry out these
investigations.

\section*{Acknowledgments}

I am grateful to my colleagues Dick Furnstahl, Carrie Halkyard, Dave
Madland, Huabin Tang, and Dirk Walecka for constructive comments and
useful discussions. This work was supported in part by the
Department of Energy under Contract No.~DE-FG02-87ER40365.

\appendix*
\section{Isovector Currents}


For completeness, we list here the expressions for the isovector
vector and axial-vector currents originating from the lagrangian
(\ref{eq:eft-lagrangian}) to all orders in the pion fields. These
expressions generalize Eqs.~(142) and (143) of Ref.~\cite{AXC}.
\begin{eqnarray}
V^\mu_a &\!\!=\!\!&
    -i\, \frac{\fpi^2}{4} \Tr [\tau_a (U \partial^\mu U^\dagger + U^\dagger
                             \partial^\mu U )]
    + \frac{1}{2}\, \Nbar \gammamu \left( \Xiaplus \right) N
               \nonumber\\[8pt]
& & \quad
    {}+ \frac{1}{2}\, g_A \Nbar \gammamu\gammafivel \left( \Xiaminus \right) N
    + i\, \frac{\kappa_\pi}{M}\, \Nbar \left[ \left( \Xiaminus \right) , a_\nu
                  \right] \sigmamunu N
                  \nonumber\\[8pt]
& & \quad
    {}+ \frac{4 \beta_\pi}{M}\, \Nbar N \Tr \left[ \left( \Xiaminus \right)
                  a^\mu \right]
    + i\, \frac{f_\rho g_\rho}{4 M} \, \Nbar \left[ \left( \Xiaplus \right) ,
                  \rho_\nu \right] \sigmamunu N
                  \nonumber\\[8pt]
& & \quad
    {}+ 2i g_{\rho\pi\pi} \frac{\fpi^2}{m^2_\rho} \, \Tr \left\{ \rho^{\mu\nu}
                  \left[ \left( \Xiaminus \right) , a_\nu \right]
             + v^{\mu\nu} \left[ \left( \Xiaplus \right) , \rho_\nu \right]
                  \right\} \nonumber\\[8pt]
& & \quad
    {}+ i \Tr \left\{ \left[ \left( \Xiaplus \right) ,
                  \rho_\nu \right] \rho^{\mu\nu} \right\} \ ,
    \label{eq:Vtotal}
\end{eqnarray}
\begin{eqnarray}
A^\mu_a &\!\!=\!\!&
    -i\, \frac{\fpi^2}{4} \Tr [\tau_a (U \partial^\mu U^\dagger - U^\dagger
                             \partial^\mu U )]
    - \frac{1}{2}\, \Nbar \gammamu \left( \Xiaminus \right) N
               \nonumber\\[8pt]
& & \quad
    {}- \frac{1}{2}\, g_A \Nbar \gammamu\gammafivel \left( \Xiaplus \right) N
    - i\, \frac{\kappa_\pi}{M}\, \Nbar \left[ \left( \Xiaplus \right) , a_\nu
                  \right] \sigmamunu N
                  \nonumber\\[8pt]
& & \quad
    {}- \frac{4 \beta_\pi}{M}\, \Nbar N \Tr \left[ \left( \Xiaplus \right)
                  a^\mu \right]
    - i\, \frac{f_\rho g_\rho}{4 M} \, \Nbar \left[ \left( \Xiaminus \right) ,
                  \rho_\nu \right] \sigmamunu N
                  \nonumber\\[8pt]
& & \quad
    {}- 2i g_{\rho\pi\pi} \frac{\fpi^2}{m^2_\rho} \, \Tr \left\{ \rho^{\mu\nu}
                  \left[ \left( \Xiaplus \right) , a_\nu \right]
             + v^{\mu\nu} \left[ \left( \Xiaminus \right) , \rho_\nu \right]
                  \right\}  \nonumber\\[8pt]
& & \quad
    {}- i \Tr \left\{ \left[ \left( \Xiaminus \right) ,
                  \rho_\nu \right] \rho^{\mu\nu} \right\} \ .
    \label{eq:Atotal}
\end{eqnarray}
Note that the sign of the first term in Eq.~(107) [and Eq.~(110)] of
Ref.~\cite{AXC} is incorrect. This propagates into an additional
$\Nbar N \pi \rho$ coupling in the axial-vector current of
Eq.~(143), namely,
\begin{equation}
A^{a\,\lambda}_{\mathrm{new}} = { {1} \over {f_{\pi}} } \,
        { {f_{\rho} g_{\rho}} \over {4M} } \,
{\epsilon}^{abc} {\epsilon}^{cdf} {\pi}^b {\rho}^d_{\nu} \, \Nbar
{\sigma}^{\lambda \nu} {\tau}^f N  \ .
\end{equation}
This additional term affects neither the charge algebra nor the
tree-level, two-nucleon amplitudes of that paper.

%

\end{document}